\documentclass{evn2004}
\usepackage{txfonts}
\begin{document}
   \title{Spacecraft Design of VSOP-2}

   \author{Y.Murata\inst{1},
           H.Hirabayashi\inst{1}
          \and
          Next Generation Space VLBI Working Group\inst{1,2}
          }

   \institute{Japan Aerospace Exploration Agency,
      Institute of Space and Astronautical Science,
      Yoshinodai 3-1-1, Sagamihara, Kanagawa, 229-8510, Japan
         \and
      National Astronomical Observatory of Japan,
      Osawa 2-21-1, Mitaka, Tokyo, 181-8588, Japan
             }

   \abstract{
As presented by Hirabayashi et al.\ in these proceedings, the
VSOP-2 mission is currently being planned. 
Various kinds of developments are being made for the misson,
and here we introduce the large antenna, fast switching scheme
using CMG, low noise receivers, gigabit data transmission, and high data
rate sampling on-board.
We are also studying the system configuration of the VSOP-2 satellite and the orbit appropriate for the expected launch vehicle, the M-V rocket. 
VSOP-2 science goals include imaging the accretion disks around
the massive blackholes in the nuclei of the active galaxies, studies of
magnetic field orientation and evolution in jets, and the highest resolution
studies of spectral line masers and mega-masers.
Here we describe the design and the development of the space antenna,
high speed sampler, and the satellite system.
   }

   \maketitle
%

\section{Description of VSOP-2 satellite}

The VSOP-2 satellite has a 10m-class antenna with low noise receivers and
a downlink data rate of 1\,Gbps.
It will detect both LCP and RCP in the 8, 22, and 43\,GHz bands.
The 22 and 43\,GHz receivers are cryogenically
cooled to reduce the system temperature to around 30\,K.
We expect a $\sim$10 times higher sensitivity than that of the VSOP mission with
these upgrades of the observing system.

Observing frequencies up to 43\,GHz will allow $\sim$10 times higher angular
resolution than that of the 5\,GHz observations of the VSOP mission.
The apogee height of 25,000\,km will allow an angular resolution of
38\,micro-arcseconds to be achieved at 43\,GHz, corresponding to around
10 Schwarzschild radii at the distance of M87.
The 10 times higher frequency band will also allow us to observe regions closer to
the AGN cores and/or the accretion disk.

Several possibilities for phase-referencing to improve sensitivity by
extending the coherence time are under study.
A comparison of the specifications of the VSOP and VSOP-2 satellites
is shown in Table~\ref{tab:spec}.

   \begin{figure}
   \centering
   \vspace{180pt}
   \includegraphics{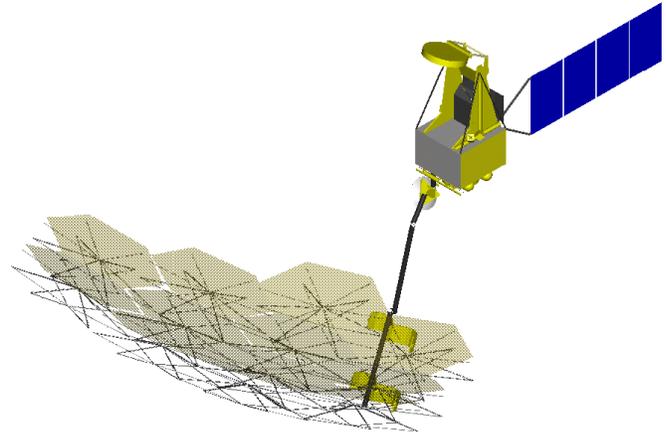}
      \caption{A schematic view of VSOP-2 satellite.
         \label{fig:vsop2sat}
         }
   \end{figure}
%

   \begin{table}
      \caption[]{Comparison of the specifications of VSOP and VSOP-2.} 
         \label{tab:spec}
     $$ 
         \begin{array}{p{0.5\linewidth}rr}
            \hline
            \noalign{\smallskip}
             items     &  VSOP & VSOP-2 \\
            \noalign{\smallskip}
            \hline
            \noalign{\smallskip}
            Antenna Diameter & 8 m & 9 m     \\
            Apogee Height & 21,500 km & 25,000 km    \\
            Orbit Period  & 6.3 h & 7.5 h \\
            Polarization & LCP & LCP \& RCP              \\
            Data Bandwidth & 128 Mbps & 1 Gbps \\
            Observing Band (GHz) & 1.6, 5, (22) & 8, 22, 43 \\
            Maximum resolution & 0.36 mas & 0.038 mas \\
            Sensitivity$^{\mathrm{a}}$ (5/8 GHz) & 158 mJy & 22 mJy \\
              \hfil          (22 GHz)  &  - & 39 mJy \\
             Phase-referencing  & & \\
             Sensitivity (1.5h integration) & - & 9.1 mJy \\
             Launch (earliest case) & 1997 & 2011 \\
            \noalign{\smallskip}
            \hline
         \end{array}
     $$ 
\begin{list}{}{}
\item[$^{\mathrm{a}}$] Assuming a VLBA antenna as the ground radio telescope.
\end{list}
   \end{table}

\section{Design of the VSOP-2 satellite}

A schematic view of the VSOP-2 satellite is shown in figure~\ref{fig:vsop2sat}.
The VSOP-2 satellite will be launched by the ISAS M-V rocket (which was also used to launch HALCA in 1997).
The orbit will have a 25,000\,km apogee height, 1,000\,km perigee height, and
a 7.5\,hour period.
The design of the satellite is highly affected by the size of the nose-faring
of the M-V rocket.
The target total weight is 910 kg.
The design of the VSOP-2 satellite is based on HALCA design, but there are
some new items we must take into account:
(1) a lighter, but larger, off-axis paraboloid antenna.
(2) the cooler, which requires more power.
VSOP-2 is expected to use a total power of about 1.8\,kW, approximately
double that of HALCA.
This means that we must take care about the management of the thermal control more carefully.
(3) the addition of 2 CMG's (control moment gyro) for fast switching observations,
GPS receivers and an accelerometer for the cm-order orbit
determination, to carry out phase-referencing observations.

\section{Developments for VSOP-2 satellite}

\subsection{Antenna Development}
\begin{figure}
   \centering
   \vspace{180pt}
   \includegraphics{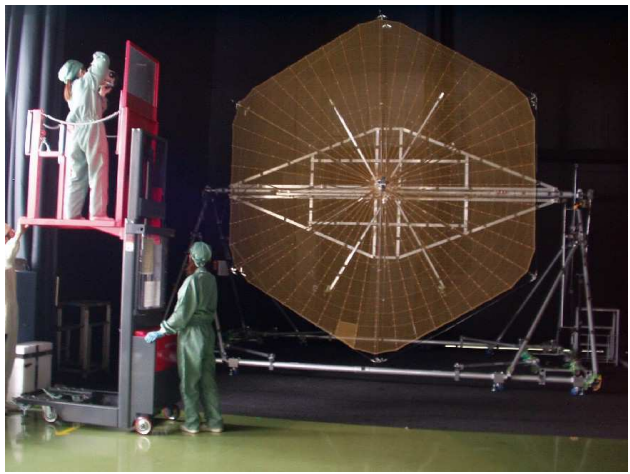}
      \caption{Measurement of the surface accuracy of a full scale model
      of a single antenna module, using a camera.
         \label{fig:vsop2ant}
         }
   \end{figure}

One of the technical challenges is the requirement placed on the surface
accuracy of the mesh antenna by the highest observing frequency of 43\,GHz
(a wavelength of 7\,mm).
A 7-segment, offset 10-m class antenna is under development.
This antenna has a lighter weight than that of the HALCA antenna,
and incorporates a tilt and focus adjustment mechanism.
The basic concept of this type of antenna was developed by NASDA (now JAXA)
for the ETS-VIII mission.
However, the planned surface accuracy of ETS-VIII is lower than that required
for the VSOP-2 mission.
We made the full scale 1 segment model (figure~\ref{fig:vsop2ant})
to evaluate how accurately the antenna surface can be adjusted.
We have determined the best way to adjust the surface of the antenna in the
Earth's gravity environment with this test.

%

\subsection{High speed sampler radiation test}
One of the problems for VSOP-2 is the on-board data sampling,
if we follow the same approach as HALCA.
We assume a giga-bit rate of data sampling to obtain the desired sensitivity.
However, we have not been able to locate any A/D sampler LSI chips qualified for use in space.
We tested a 10\,Gbps, 1\,bit sampler and a 1:16 demultiplexer, in an effort to
find a possible solution for the on-board high speed sampler.
We carried out a 1000\,krad total dose test and a single event test,
and confirmed we can use those LSI's on the VSOP-2 orbit.

\subsection{Giga-bit data link}
A giga-bit data transmission requires more than a 1\,GHz bandwidth frequency
allocation.
The possible band, based on frequency allocation regulations, will be 37--38\,GHz.
The uplink frequency will be moved to 40\,GHz.
We studied the link budget, and found that the condition is more severe
than that of HALCA Ku link, but it is still possible.
We are developing the modulator and demodulator for the giga-bit data transmission.
We are testing the OFDM (Orthogonal Frequency Division Multiplex) method
with 8 channel QPSK modulation.

\subsection{Cooled frontend.}
A 43\,GHz GaAs MMIC cooled receiver is being developed using the commercial MMIC LSI
for the room temperature use.
Currently, the noise temperature is about 40\,K at a 30\,K physical
temperature, which is not the lowest possible, but is sufficient.
We also plan to test an InP HEMT amplifier to achieve a lower noise temperature, and
adjust the low power supply to make the cooling system easier.
We also developing the cooled frontend system for the 22 and 43\,GHz bands.
We use the Stirling cycle refrigerator, which has been developed for ASTRO-F
(infrared astronomy) mission by JAXA, and confirmed that we can achieve a physical
temperature of less than 30\,K.

\subsection{The phase-referencing capability}
Phase referencing observations are a powerful method to determine accurate
relative positions of radio sources, and longer integration times for weaker
sources.
High speed switching between sources on a less than 1 minute cycle,
and orbit determination accuracy of less than 3\,cm are required.
We are now studying the possibility of fast switching using
2 control moment gyro's (CMG) which are a higher torque actuator,
with the combination of 4 reaction wheels which are used for the normal
attitude control.
We have succeeded to make the fast switching with a 1 minute period for a 3.5$^\circ$ separation. The possibility of obtaining a higher orbit accuracy using
an on-board GPS system and accelerometer is also being studied (Wu and Bar-Sever 2001).

\section{Conclusions}

We are now preparing the VSOP-2 mission proposal for submission to ISAS/JAXA. There are some of technically challenging items, but we have made good progress in developing solutions for them. 

\begin{acknowledgements}
Finally, we gratefully acknowledge all the collaborators who have joined discussions about the VSOP-2 mission. This mission is planned based on discussions with researchers from JPL, NRAO, CfA (US), DRAO, SGL (Canada), ATNF (Australia), JIVE (Europe), Univ.\ of Ibaraki, Hosei Univ., Yamaguchi Univ., NAOJ, and ISAS (Japan)\end{acknowledgements}

\end{document}